# EGovernment Stage Model: Evaluating the Rate of Web Development Progress of Government Websites in Saudi Arabia

Osama Alfarraj, Steve Drew
ICT School, Griffith University,
Brisbane, Australia

Rayed Abdullah AlGhamdi
King Abdulaziz University,
Jeddah, Saudi Arabia

*Abstract –* **This paper contributes to the issue of eGovernment implementation in Saudi Arabia by discussing the current situation of ministry websites. It evaluates the rate of web development progress of vital government websites in Saudi Arabia using the eGovernment stage model. In 2010, Saudi Arabia ranked 58[th] in the world and 4[th] in the Gulf region in eGovernment readiness according to United Nations reports. In particular, Saudi Arabia has ranked 75[th] worldwide for its online service index and its components compared to the neighbouring Gulf country of Bahrain, which was ranked 8[th] for the same index. While this is still modest in relation to the Saudi government's expectation concerning its vision for eGovernment implementation for 2010, and the results achieved by the neighbouring Gulf countries such as Bahrain and the United Arab Emirates on the eGovernment index, the Saudi government has endeavoured to meet the public needs concerning eGovernment and carry out the implementation of eGovernment properly. Governments may heed the importance of actively launching official government websites – the focus of this study – as the main portals for delivering their online services to all the different categories of eGovernment (including G2C, G2B, and G2G). However, certain Saudi ministries have not given due attention to this vital issue. This is evidenced by the fact that some of their websites are not fully developed or do not yet exist, which clearly impedes that particular ministry from appropriately delivering eServices.**

## I. INTRODUCTION

In applying the concept of eGovernment in Saudi Arabia, the supreme Royal Decree number 7/B/33181 of 7 September 2003 was established to transform Saudi society into an information society by initiating and supporting new strategies and efforts to facilitate the electronic delivery of government services [1, 2]. However, this decree was not launched until 2005 [2].

The vital initiatives for implementing eGovernment should be followed and given due attention in order to enhance the delivery and use of eGovernment services. One of these initiatives is the active launch of official government websites as the main portals for delivering online services to all of the different categories of eGovernment, including G2C, G2B, and G2G. However, some Saudi ministries do not seem to be paying adequate attention to this issue, as some of their websites are not well developed or do not yet exist, which absolutely impedes that particular ministry from appropriately delivering eServices.

In the study, 'E-Government in Saudi Arabia: Can it overcome its challenges?' conducted by Sahraoui et al. [2], the researchers indicated that only 13 out of 22 Saudi ministries have an online presence. This represents 60% of the ministries. Conducted in 2006, this study was based on a survey as well as online browsing and the accessing of over 25 government websites to evaluate these websites [2]. Moreover, the researchers indicated that the number of Saudi ministries with an online presence has remained the same since the study that had been done by Abanumy et al in 2003 [2], who also found that 'only 13 ministries had online presence, while 8 did not' (as illustrated in Table I below), and that none of these websites were accessible to disabled people as cited in [2, 3].

TABLE I.     ONLINE SURVEY FOR SAUDI GOVERNMENT WEBSITES CONDUCTED BY ABANUMY AND MAYHEW IN 2003

| Stage reached | Assessment elements | Number | % |
|---|---|---|---|
| No presence | No official website available | 8 | 38% |
| Emerging presence | e.g. agency name, agency phone number, address, operating hours, general frequently asked question | 0 | 0% |
| Enhanced presence | e.g. organisational news, publication, online policy (security, privacy) | 3 | 14% |
| Interactive presence | e.g. officials' e-mail, post comment online, simple two-way communication, download organization's forms | 10 | 48% |
| Transactional presence | e.g. e-form, e-payment | 0 | 0% |
| Seamless | Full integration across organisation | 0 | 0% |

Source: Adapted from Abanumy et al. [3].

According to [4], 'six out of every ten government departments with an Internet connection have their own website. Less than 10% of the websites are hosted only in English. The majority of websites are hosted in both English and Arabic (52%) a shift from Arabic only in 2007' (p. 24).

Although developing eGovernment web portals requires tremendous effort and resources, including human resources, software and hardware, there seems to be a delay in building and launching the government web portals, even though each government agency is in charge of developing its own website and has full responsibility for doing so [5, 6]. For example, one of the main government websites which still lacks an online presence is the Ministry of Hajj (pilgrimage). Approximately two million people from all over the world come at a certain time every year to perform Hajj in Makkah,

Saudi Arabia. Such an influx of visitors requires comprehensive information and services to facilitate their travel, such as applying for a Hajj Visa or other related services. The availability a website that would provide online services and facilitate information sharing would benefit all Muslims worldwide, as well as the people of Saudi Arabia. While another website, www.hajinformation.com, does provide information on all issues related to Hajj, it is not an official website for the ministry and lacks online services. The failure to develop such essential websites for all citizens, residents and businesses is delaying the implementation of eGovernment in Saudi Arabia as a part of the eGovernment project.

## II. AIM AND SIGNIFICANCE OF THIS STUDY

Despite the emphasis on the concept of eGovernment in the literature, there is still a lack of research which evaluates the progress of government websites, specifically in Saudi Arabia, to show where these websites stand in terms of their readiness to deliver eServices. According to [7], Saudi Arabia was ranked as number 58 worldwide for eGovernment readiness index in 2010 and 4th among the Gulf countries. This position is far from the expectation for 2010, as the Saudi government had predetermined that 'by the end of 2010, everyone in the Kingdom will be able to enjoy from anywhere and at any time – world class government services offered in a seamless, user friendly and secure way by utilizing a variety of electronic means'. As it is now 2011, the timetable for the eGovernment program set by the Saudi government is not being achieved as expected in light of what has been done so far and as indicated in the literature. Thus, this study was conducted to show the level of web readiness of Saudi's government websites, which has played a significant role in delaying the delivery of eServices.

## III. RESEARCH METHODOLOGY

The aim of this study is to evaluate the web development of government websites in Saudi Arabia and then repeat the same evaluation for Bahraini government websites to reveal the level of readiness of Saudi government websites in comparison to Bahrain. In addition to helping determine the current status of the Saudi government websites, this study can also determine whether the websites are ready to enter the transactional stage as online service providers. As the study by Sahraoui et al. [2] to rate Saudi websites was conducted in 2006 and no current evaluation study was available in the literature at the time of writing, Saudi ministry websites need to be re-evaluated to note any differences in the subsequent four years (from 2006 to 2010). This will primarily be done by reviewing the relevant published literature and using the eGovernment stage model adopted by [2, 8, 9].

The researchers browsed and visited the same websites of Saudi ministries that Sahraoui et al. [2] evaluated as well as some others. Moreover, the same government websites were also selected in the Bahrain context for the evaluation. The researchers used a checklist (assessment elements) that helps to determine the proper stage of each government website selected, which has been attached as an appendix to this study.

## IV. EGOVERNMENT STAGE MODEL

As shown in Figures 1 and 2 below, the eGovernment stage model has five main stages: (I) Emerging presence; (II) Enhanced presence; (III) Interactive presence; (IV) Transactional presence; and (V) Seamless or connected [10, 11].

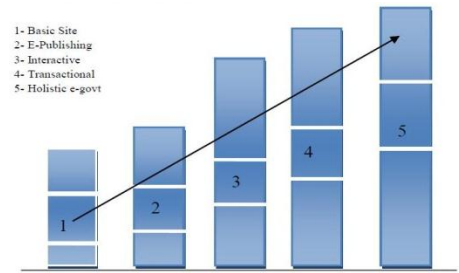

Figure 1. Stages of the eGovernment model (Adopted from [12, 2]).

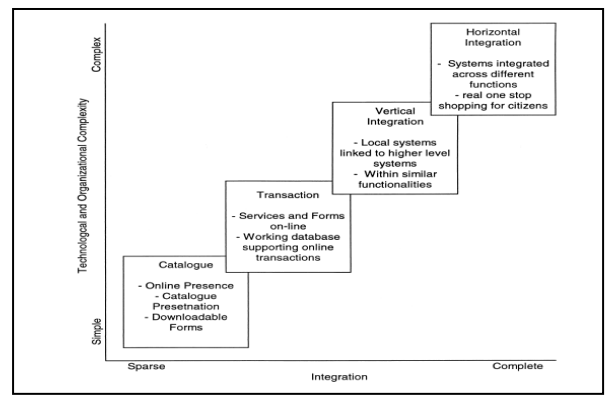

Figure 2. Dimensions and Stages of the eGovernment model (Adopted from [11]).

The stages of eGovernment are further clarified below.

*1) Stage I is referred to as emerging. To fall into this stage of eGovernment online presence, there should be an official website for the country containing information about it and at there must be links for the country's ministries and departments, such as health, education and so on [13].*

*2) Stage II is the enhanced stage wherein the government provides more information to citizens on public policy and the government as well as other information such as reports and regulations, all of which is easily and continuously accessible through archives [10].*

*3) Stage III is the interactive stage. In this stage the government provides downloadable forms for other services in order to enhance the ease and convenience of the service requester. Simple two-way communication with the ability to post comments online is also offered [13].*

*4) Stage IV is the transactional stage, which is when the government has started to provide online services and allows citizens to access these services 24/7 in order to represent G2C interactions. Examples of these services are applications for ID cards and online license renewals [10].*

*5) In Stage V, the connected stage, governments activate back offices; that is, they have transformed themselves into an online entity that meets their citizens' needs and can respond to their citizens in easy and modern ways. Thus, it represents the most developed level of online government initiatives and has the following characteristics:*

*a) Horizontal connections (among government agencies)*

*b) Vertical connections (central and local government agencies)*

*c) Infrastructure connections (interoperability issues)*

*d) Connections between governments and citizens*

*e) Connections among stakeholders (government, private sector, academic institutions, NGOs and civil society) [10].*

## V. EVALUATION OF SAUDI AND BAHRAINI GOVERNMENT WEBSITES USING THE EGOVERNMENT STAGE MODEL

### A. Saudi Context

Only a few studies have been conducted to evaluate the rate of developmental progress of government websites in Saudi Arabia using the eGovernment stage model. The study done by [2] evaluated the Saudi ministries' web portals in accordance with the eGovernment stage model. The researchers of this study browsed and visited some of the main Saudi government websites to determine at which stage each one of these government websites was for the purpose of showing their readiness. As the researchers state, 'they are mostly situated between stage II and III of the e-government stage model, hence not yet fully transactional' [2]. As only 13 out 22 Saudi ministries had an online presence, with the majority of these websites being placed between stages II and III, none were providing online services. Table II depicts the majority of Saudi government websites that were placed between stages II and III. These websites are considered to be only information providers rather than service providers, with the exception of Saudi Telecom, a government organization which was placed at stage IV as a service provider.

As the study by [2] was done in 2006 and no current evaluation study was available in the literature at the time of writing, the evaluation of the Saudi ministries needs to be done again to note the differences in the next four years (from 2006 to 2010) and to show the current level of these web portals. Thus, the same assessment elements were employed by Sahraoui and his colleagues in 2006 were used in this study. The researchers browsed and visited the same Saudi ministries' websites as well as some others (see Table III) and were evaluated according to the eGovernment stage model offered by United Nations to determine whether or not these official web portals have started delivering their eServices to their citizens, residents, and businesses. This links with the above discussion on whether the Saudi eGovernment has implemented and delivered its eServices well in the specified time set in 2005 to the 'Yesser' program, which stated: 'By the end of 2010, everyone in the Kingdom will be able to enjoy – from anywhere and at any time – world class government services offered in a seamless, user friendly and secure way by utilizing a variety of electronic means' [14, 2, 15, 16]. As seen in Table III below, 28 official Saudi government websites of various ministries were browsed and visited via the online survey.

The online survey examined the following criteria: whether the ministry has an online presence, information, downloadable applications, online applications, transaction inquiry, transaction online (services online), and whether an English version is available for the same website and other checklist criteria (see appendix), as carried out by [2] and previously by [9, 8].

TABLE II.     ONLINE SURVEY OF SAUDI GOVERNMENT WEBSITES CONDUCTED BY SAHRAOUI ET AL. ON 6 APRIL 2006

| | Authority | Application Download | Application Online | Transaction Inquiry | Transaction Online | English Version | Stage | URL |
|---|---|---|---|---|---|---|---|---|
| 1 | The E-government Program | No | Yes | Yes | No | No | ? | www.gov.sa |
| 2 | Communication and Information Technology Commission | Yes | Yes | Yes | Yes | No | II | www.citc.gov.sa |
| 3 | Saudi Telecom | Yes | Yes | Yes | Yes | Yes | IV | www.stc.com.sa |
| 4 | Ministry of Agriculture | Yes | No | Yes | No | No | III | www.agrwat.gov.sa |
| 5 | Ministry of Civil Service | Yes | No | Yes | No | No | III | www.mcs.gov.sa |
| 6 | Ministry of Commerce and Industry | Yes | No | Yes | No | No | III | www.commerce.gov.sa |
| 7 | Council of Saudi Chambers of Commerce and Industry | Yes | Yes | Yes | * | Yes | III | www.saudichambers.org.sa |
| 8 | Ministry of Defense and Aviation | Yes | No | Yes | No | Yes | III | www.gaca.gov.sa |
| 9 | Ministry of Water and Electricity | Yes | No | Yes | No | No | III | www.mow.gov.sa |
| 10 | Saudi Ports Authority | No | No | No | No | No | III | www.ports.gov.sa |
| 11 | Ministry of Interior Passport Authority | Yes | No | Yes | No | No | II | www.passport.gov.sa |
| 12 | Ministry of Health | Yes | No | Yes | No | No | III | www.moh.org.sa |
| 13 | Ministry of Education | Yes | No | Yes | No | No | III | www.moe.gov.sa |
| 14 | Royal Embassy of Saudi Arabia - London | No | No | Yes | No | Yes | III | www.mofa.gov.sa |
| 16 | Ministry of Hajj | No | No | No | No | No | III | www.hajinformation.com |
| 16 | Ministry of Higher Education | Yes | No | Yes | No | No | III | www.mohe.gov.sa |
| 17 | Ministry of Communications and Information Technology (IT) | No | No | No | No | Yes | III | www.mcit.gov.sa |

This online survey revealed that only two of the selected ministries (Ministry of Hajj and General Presidency of Youth Welfare) still have no online presence, while 26 have an online presence, which shows improvement compared to the online survey done by [2].

One of these two ministries lacking an online presence is the Ministry of Hajj (pilgrimages), which is a very important ministry that benefits not only the residents, citizens, and businesses of Saudi Arabia, but all Muslims worldwide as well.

TABLE III. ONLINE SURVEY FOR SAUDI GOVERNMENT WEBSITES CONDUCTED BY THE RESEARCHERS OF THIS STUDY IN SEPTEMBER 2010

| No. | Authority | Has presence | Information | Application Download | Application Online | Transaction Inquiry | Transaction Online | English Version | Stage | URL |
|---|---|---|---|---|---|---|---|---|---|---|
| 1 | Saudi eGovernment national portal | Yes | Yes | Yes | Yes | Yes | No | Yes | III | http://www.saudi.gov.sa |
| 2 | Communication and Information Technology Commission | Yes | Yes | Yes | Yes | Yes | No | Yes | III | http://www.citc.gov.sa |
| 3 | Saudi Telecom | Yes | Yes | Yes | Yes | Yes | Yes | Yes | IV | http://www.stc.com.sa |
| 4 | Ministry of Agriculture | Yes | Yes | Yes | No | Yes | No | No | III | http://www.moa.gov.sa |
| 5 | Ministry of Civil Service | Yes | Yes | Yes | Yes | Yes | Yes | No | IV | http://www.mcs.gov.sa |
| 6 | Ministry of Commerce and Industry | Yes | Yes | Yes | Yes | Yes | Yes | Yes | IV | http://www.mci.gov.sa |
| 7 | Council of Saudi Chambers of Commerce and Industry | Yes | Yes | Yes | Yes | Yes | Yes | Yes | IV | http://www.saudichambers.org.sa |
| 8 | Ministry of Defense and Aviation | Yes | Yes | Yes | Yes | Yes | Yes | Yes | IV | ttp://www.gaca.gov.sa |
| 9 | Ministry of Water and Electricity | Yes | Yes | Yes | Yes | Yes | Yes | Yes | IV | http://www.mowe.gov.sa/ |
| 10 | Saudi Ports Authority | Yes | Yes | No | No | No | No | Yes | II | http://www.ports.gov.sa |
| 11 | Ministry of Interior (Passport Authority) | Yes | Yes | Yes | Yes | Yes | No | No | III | http://www.gdp.gov.sa |
| 12 | Ministry of Health | Yes | Yes | No | Yes | Yes | No | Yes | IV | http://www.moh.gov.sa |
| 13 | Ministry of Education | Yes | Yes | Yes | Yes | Yes | No | Yes | III | http://www.moe.gov.sa |
| 14 | Ministry of Hajj | No | Yes | No | No | No | No | Yes | III | http://www.hajjinformation.com/ |
| 15 | Ministry of Foreign Affairs | Yes | Yes | Yes | Yes | Yes | Yes | Yes | IV | http://www.mofa.gov.sa |
| 16 | Ministry of Higher Education | Yes | Yes | Yes | Yes | Yes | Yes | Yes | V | www.mohe.gov.sa |
| 17 | Ministry of Communications and Information Technology (IT) | Yes | Yes | Yes | Yes | Yes | No | Yes | III | http://www.mcit.gov.sa |
| 18 | Ministry of Finance | Yes | Yes | Yes | Yes | Yes | No | Yes | III | http://www.mof.gov.sa |
| 19 | Ministry of Justice | Yes | Yes | Yes | No | No | No | No | III | http://www.moj.gov.sa |
| 20 | Ministry of Labor | Yes | Yes | Yes | Yes | Yes | Yes | Yes | IV | http://www.mol.gov.sa |
| 21 | Ministry of Municipalities and Rural Affairs | Yes | Yes | Yes | Yes | Yes | Yes | No | IV | http://www.momra.gov.sa/ |
| 22 | Ministry of Petroleum and Mineral Resources | Yes | Yes | No | No | No | No | Yes | II | http://www.mopm.gov.sa |
| 23 | Ministry of Economy and Planning | Yes | Yes | Yes | Yes | Yes | No | Yes | III | http://www.mep.gov.sa |
| 24 | Ministry of Islamic Affairs, Endowment, Dawa and Guidance | Yes | Yes | Yes | Yes | Yes | Yes | Yes | IV | http://www.moia.gov.sa |
| 25 | Ministry of Social Affairs | Yes | Yes | Yes | Yes | No | No | No | III | http://mosa.gov.sa |
| 26 | Ministry of Transport | Yes | Yes | Yes | Yes | Yes | No | Yes | III | http://www.mot.gov.sa |
| 27 | Ministry of Interior | Yes | Yes | Yes | Yes | Yes | Yes | Yes | IV | http://www.moi.gov.sa |
| 28 | General Presidency of Youth Welfare | Yes | * | * | * | * | * | * | * | http://www.gpyw.gov.sa |

The website of Ministry of Hajj, which was provided and examined in the table above, is an unofficial website and does not officially belong to the Ministry of Hajj.

Thus, specific attention should be given to the development of this Ministry's web portal. Moreover, the online survey revealed that only two government websites, namely, the Saudi Ports Authority and the Ministry of Petroleum and Mineral Resources, are designated as falling under Stage II (enhanced presence) of the eGovernment stage model, which means that they are only providing basic items that are not even in downloadable forms. Therefore, these two government websites are at a very low level and need to be developed as both are essential ministries. Table IV below summarises the results of the online survey accroding the number of government organizations in each stage.

As seen in Table IV, the online survey indicated that 11 government ministry websites are currently at stage III (Interactive presence), which means that these ministries do not yet provide online services.

TABLE IV. THE NUMBER OF SAUDI GOVERNMENT WEBSITES IN EACH STAGE, AS SURVEYED IN SEPTEMBER 2010

| Stage No. | Stage reached | Assessment elements | Number of Saudi government ministries |
|---|---|---|---|
| | No presence | No official websites available | 2 |
| I | Emerging presence | **e.g.** agency name, agency phone number, address, operating hours, general frequently asked questions | - |
| II | Enhanced presence | **e.g.** organisational news, publication, online policy (security, privacy) | 2 |
| III | Interactive presence | **e.g.** officials' e-mail addresses, ability to post comments online, simple two-way communication, can download the organisation's forms | 11 |
| IV | Transactional presence | **e.g.** e-form, e-payment and some query services | 12 |
| V | Seamless | Full integration across the organisation | 1 |

On the other hand, the online survey found that 12 Saudi ministry websites are currently at stage IV (transactional presence), demonstrating that these ministry websites are considered to be online services providers. The majority of these websites have basic online services such as online query services and eForms. Additionally, the online survey found that only one ministry is currently at stage at stage V (seamless), namely, the Ministry of Higher Education. Consequently, none of the government websites, except for the Ministry of Higher Education, achieved full integration or even a high performance at the level of transactional presence by the end of 2010 as recommended by the Yesser program. In comparison with informational websites, transactional websites usually receive high scores for the United Nations web index for eGovernment readiness [10]. This is why the 'survey ranked the UAE in $5^{th}$ position in terms of transactional services, just behind developed countries like Sweden, Denmark, Norway and the US', as shown in Figure 3 below [17, 18]. The transactional stage was defined by [19] as one 'in which citizens will be able to conduct business online with governments'.

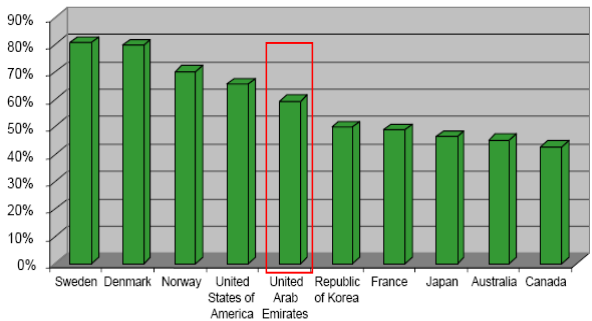

Figure 3.    Transactional services: Top 10 countries in 2008 [10].

### B.  Bahraini Context

Bahrain is a neighbouring country to Saudi Arabia and was used in this study as a country of comparison. Table V provides a comparison of attributes between Saudi Arabia and Bahrain in relation to the readiness of their government websites to deliver eServices.

The same online evaluation survey was conducted evaluate and show the rate of web development progress for Bahraini government websites compared to those evaluated in the Saudi Arabian context (as show in table VI). This online evaluation survey has revealed that all 23 selected ministries have an online presence and that none of the selected ministries are at Stage II (enhanced presence) as mere information providers. Furthermore, the online evaluation found that only six government websites are designated as being in Stage III (Interactive presence) of the eGovernment stage model, indicating that none of these ministries have provided online services yet.

On the other hand, the online survey found that eight Bahraini government websites are currently at Stage IV (transactional presence), demonstrating that these ministry websites are considered to be online service providers with e-forms and payment options. Surprisingly, the online survey revealed that nine ministries are currently at Stage V

(seamless). Consequently, the majority of the selected ministries for Bahrain (which are similar to those selected for Saudi Arabia) are currently placed in the highest stage of the eGovernment model and have indeed transformed themselves into an online entity that meets its citizens' needs and responds to its citizens in easy and developed ways. Thus, this represents the most developed level of online government initiatives. Table VII summarises the results of the online evaluation survey according to the number of government organizations in each stage.

TABLE V.    THE ATTRIBUTES OF SELECTING BAHRAIN AS A COUNTRY OF COMPARISON

| Country Attribute | Saudi Arabia | Bahrain |
|---|---|---|
| Culture | • Islam is practised by the majority of people and is part of many aspects of life<br>• The family/tribal structure is one of the cultural aspects used to identify social structure and individual identity<br>• Conservative dress for both men and women is compulsory | • Islam is practised by the all people and it is part of many aspects of life<br>• The family/tribal structure is one of the cultural aspects used to identify social position and individual identity<br>• Conservative dress for both men and women is not compulsory, but this does not mean that it should be too free. It should be acceptable and not offend public decency |
| Government and the country | • Monarchy<br>• Constitution governed in accordance with Islamic Law<br>• Legal system based on Sharia law (Islamic law)<br>• Unification of the Kingdom as a public holiday alongside with *Eid ul-Fitr* (festivity of conclusion of the fast) and *Eid al-Adha* (Festival of Sacrifice) | • Monarchy<br>• Legal system based on Islamic law and English common law<br>• There are several public holidays which include New Year's Day. Mouloud (Birth of the Prophet), Eid al-Fitr (End of Ramadan), Eid al-Adha (Feast of the Sacrifice), National Day (two days), Labour day, Accession Day, Arafat day, Al-Hijrah (Islamic New Year), Ashoura which is the tenth day of Muharram in the Islamic calendar |
| People | • Arabs | • Arabs |
| Religion | • Islam | • Islam |
| Language used | • Arabic is the official language<br>• English is a commercial language | • Arabic is the official language<br>• English is a commercial language |
| Geography | • Middle East, bordering the Persian Gulf and the Red Sea | • Middle East, archipelago in the Persian Gulf, east of Saudi Arabia |
| Economy | • Oil-based economy as a major economy<br>• Main exports: oil, gas, cereals | • It is well diversified and a home for multinational firms in the Gulf region<br>• Planning depends heavily on oil<br>• Main exports: petroleum and petroleum products and aluminum |
| UN eGovernment readiness as 2010 | • Fourth ranking amongst the Gulf countries | • First ranking amongst the Gulf countries |
| Gulf Cooperation Council (GCC) | • Saudi Arabia is a member of the Gulf Co-operation Council | • Bahrain is a member of the Gulf Co-operation Council |

TABLE VI.  SURVEY FOR BAHRAINI GOVERNMENT WEBSITES CONDUCTED BY RESEARCHERS IN SEPTEMBER 2010

| No. | Authority | Application Download | Application Online | Transaction Inquiry | Transaction Online | English Version | Stage | URL |
|---|---|---|---|---|---|---|---|---|
| 1 | Bahraini eGovernment national portal | Yes | Yes | Yes | Yes | Yes | V | http://www.e.gov.bh/pubportal |
| 2 | Telecommunication regulatory authority | Yes | Yes | Yes | Yes | Yes | IV | http://www.tra.org.bh/ |
| 3 | Bahraini Telecom | Yes | Yes | Yes | Yes | Yes | IV | http://www.batelco.com/portal/ |
| 4 | Ministry of Municipalities and Agriculture Affairs | Yes | Yes | Yes | Yes | Yes | IV | http://websrv.municipality.gov.bh/mun/index_en.jsp |
| 5 | Ministry of Civil Service | Yes | Yes | Yes | No | No | III | http://www.csb.gov.bh/csb/wcms/ar/home/ |
| 6 | Ministry of Industry and Commerce | Yes | Yes | Yes | Yes | Yes | V | http://www.moic.gov.bh/moic/en |
| 7 | Bahrain chamber of commerce and industry | Yes | Yes | Yes | No | Yes | III | http://www.bahrainchamber.org.bh |
| 8 | Bahrain defence force | Yes | Yes | Yes | No | Yes | III | http://www.bdf.gov.bh/ar/default.asp |
| 9 | Electricity and Water Authority | Yes | Yes | Yes | Yes | Yes | V | http://www.mew.gov.bh |
| 10 | General organisation of sea ports, Bahrain | Yes | Yes | Yes | Yes | Yes | IV | http://www.gop.bh/index.asp |
| 11 | Ministry of Interior Passport Authority | Yes | Yes | Yes | Yes | Yes | IV | http://www.gdnpr.gov.bh/ |
| 12 | Ministry of Health | Yes | Yes | Yes | Yes | Yes | V | http://www.moh.gov.bh |
| 13 | Ministry of Education | Yes | Yes | Yes | Yes | Yes | V | http://www.moe.gov.bh |
| 14 | Ministry of Foreign Affairs | Yes | Yes | Yes | No | Yes | III | http://www.mofa.gov.bh/ |
| 15 | Ministry of Finance | Yes | Yes | Yes | Yes | Yes | V | http://www.mof.gov.bh |
| 16 | Ministry of Justice & Islamic affairs | Yes | Yes | Yes | Yes | Yes | V | http://www.moj.gov.bh |
| 17 | Ministry of Labor | Yes | Yes | Yes | Yes | Yes | IV | http://eservices.mol.gov.bh |
| 18 | National oil & gas authority | Yes | Yes | Yes | Yes | Yes | IV | http://www.noga.gov.bh |
| 19 | Ministry of Works | Yes | Yes | Yes | Yes | Yes | IV | http://www.works.gov.bh |
| 20 | Ministry of Social development | Yes | Yes | Yes | Yes | Yes | V | http://www.social.gov.bh |
| 21 | Ministry of Transport | Yes | Yes | Yes | No | Yes | III | http://www.transportation.gov.bh/ |
| 22 | Ministry of Interior | Yes | Yes | Yes | Yes | Yes | V | http://www.interior.gov.bh |
| 23 | General organization for youth and sports | Yes | Yes | Yes | No | Yes | III | http://www.goys.gov.bh/ |

TABLE VII.  THE NUMBER OF BAHRAINI GOVERNMENT WEBSITES IN EACH STAGE, AS SURVEYED IN SEPTEMBER 2010

| Stage No. | Stage reached | Assessment elements | The number of Bahraini government ministries |
|---|---|---|---|
| | No presence | No official websites available | 0 |
| I | Emerging presence | e.g. agency name, agency phone number, address, operating hours, general frequently asked questions | - |
| II | Enhanced presence | e.g. organisational news, publication, online policy (security, privacy) | 0 |
| III | Interactive presence | e.g. officials' e-mail, ability to post comments online, simple two-way communication, can download the organisation's forms | 6 |
| IV | Transactional presence | e.g. e-form, e-payment and some query services | 8 |
| V | Seamless | Full integration across the organisation | 9 |

According to [7], 'what is noteworthy is that some developing countries have begun to catch up with higher-income countries despite these challenges. Bahrain (0.7363), for example, has made significant strides in the two years since the previous survey, moving up in the rankings to 13th place in 2010 from 42nd place in 2008'. Therefore, Bahrain has already attained excellent results in respect to eGovernment readiness, having already ranked as number 13 worldwide (as illustrated in table VIII), where it ranked first amongst the Gulf countries as well as the region and third in Asia.

TABLE VIII.  TOP 20 COUNTRIES IN EGOVERNMENT DEVELOPMENT

| Rank | Country | E-government development index value | Rank | Country | E-government development index value |
|---|---|---|---|---|---|
| 1 | Republic of Korea | 0.8785 | 11 | Singapore | 0.7476 |
| 2 | United States | 0.8510 | 12 | Sweden | 0.7474 |
| 3 | Canada | 0.8448 | 13 | Bahrain | 0.7363 |
| 4 | United Kingdom | 0.8147 | 14 | New Zealand | 0.7311 |
| 5 | Netherlands | 0.8097 | 15 | Germany | 0.7309 |
| 6 | Norway | 0.8020 | 16 | Belgium | 0.7225 |
| 7 | Denmark | 0.7872 | 17 | Japan | 0.7152 |
| 8 | Australia | 0.7863 | 18 | Switzerland | 0.7136 |
| 9 | Spain | 0.7516 | 19 | Finland | 0.6967 |
| 10 | France | 0.7510 | 20 | Estonia | 0.6965 |

Source: Adapted from [30].

In fact, Bahrain ranked 8th place in eGovernment development in 2010 as demonstrated in Table IX below, while in the same context, Saudi Arabia ranked 75th [7].

According to [7], 'a country's strength in online service provision correlates positively with its use of new technology such as the emerging tools for social networking' (p. 76).

This reflects the growth and the high level of online services provided by the country with advanced Web 2.0 tools such as online discussion forums, live chat and online polls on government portals and websites. Such facilities can assist in getting citizens involved in government decision making.

TABLE IX.    TOP 20 COUNTRIES IN ONLINE SERVICE DEVELOPMENT

| Rank | Country | Online service index value | Rank | Country | Online service index value |
|------|---------|----------------------------|------|---------|----------------------------|
| 1 | Republic of Korea | 1.0000 | 11 | France | 0.6825 |
| 2 | United States | 0.9365 | 12 | Netherlands | 0.6794 |
| 3 | Canada | 0.8825 | 13 | Denmark | 0.6730 |
| 4 | United Kingdom | 0.7746 | 14 | Japan | 0.6730 |
| 5 | Australia | 0.7651 | 15 | New Zealand | 0.6381 |
| 6 | Spain | 0.7651 | 16 | Malaysia | 0.6317 |
| 7 | Norway | 0.7365 | 17 | Belgium | 0.6254 |
| 8 | Bahrain | 0.7302 | 18 | Chile | 0.6095 |
| 9 | Colombia | 0.7111 | 19 | Israel | 0.5841 |
| 10 | Singapore | 0.6857 | 20 | Mongolia | 0.5556 |

Online presence for any government website has several stages which should be gone through to maintain a high level of services provided, as mentioned previously in the eGovernment stage model section.

While attaining transactional presence is a bit challenging, it is not impossible. However, 'only a few countries are able to offer many transactional services online at this time' [7]. Bahrain is one country that has already reached this stage (as illustrated in Figure 4) and offers a wide range of integrated transactional e-services by having comprehensive back office integration systems and advanced networks. As these systems are extremely secure, they allow citizens to operate e-services with confidence [7]. This is missing in the Saudi context as the online services provided by government websites are poor and lack quality.

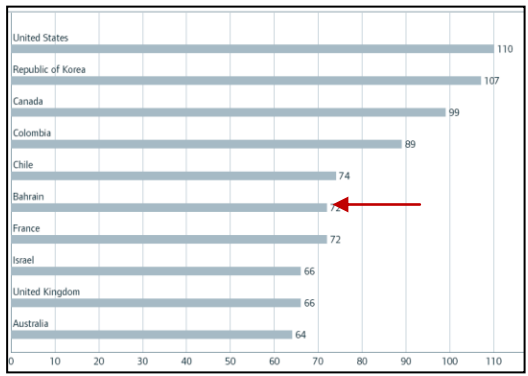

Figure 4.    Selected countries with high transactional presence scores
(Adapted from [7]).

Moreover, Bahrain has achieved a high ranking with respect to the e-participation index, which is also a part of the eGovernment readiness index for the whole country (this involves: 1. the web measure index; 2. the telecommunication infrastructure index; 3. the human capital index 4; and e-participation index) as this index can determine the level of eGovernment readiness for each country compared to others in the world. The goal of e-participation is to 'improve the citizen's access to information and public services; and participation in public decision-making' [13].

Thus, it has three parts, which are e-information, e-consultation and e-decision making, which have to be examined to determine the level of e-participation. This index easily reflects the level of engagement on the part of citizens/residents and businesses in the decision making of a particular government. Bahrain stands out among the Gulf countries as it is ranked in 11th place (as shown in Table X below) internationally and stands at 4th place in respect to the quality of e-participation websites worldwide (see Table XI). Conversely, Saudi Arabia is ranked in 102nd place with an e-participation index score (0.1000) that is considered to be very low [7]. Furthermore, Saudi government websites lack in the deployment and utilization of new technologies of Web 2.0, such as forums and web-based collaborative technologies that can effectively connect the public with the government in easy and efficient ways as is being done in Bahrain.

TABLE X.    TOP 20 COUNTRIES IN THE E-PARTICIPATION INDEX FOR 2010

| Rank | Country | 2010 e-participation index value | 2010 rank | 2008 rank | Change +/(-) |
|------|---------|----------------------------------|-----------|-----------|--------------|
| 1 | Republic of Korea | 1.0000 | 1 | 2 | 1 |
| 2 | Australia | 0.9143 | 2 | 5 | 3 |
| 3 | Spain | 0.8286 | 3 | 34 | 31 |
| 4 | New Zealand | 0.7714 | 4 | 6 | 2 |
| 5 | United Kingdom | 0.7714 | 4 | 25 | 21 |
| 6 | Japan | 0.7571 | 6 | 11 | 5 |
| 7 | United States | 0.7571 | 6 | 1 | (5) |
| 8 | Canada | 0.7286 | 8 | 11 | 3 |
| 9 | Estonia | 0.6857 | 9 | 8 | (1) |
| 9 | Singapore | 0.6857 | 9 | 10 | 1 |
| 11 | Bahrain | 0.6714 | 11 | 36 | 25 |
| 12 | Malaysia | 0.6571 | 12 | 41 | 29 |
| 13 | Denmark | 0.6429 | 13 | 3 | (10) |
| 14 | Germany | 0.6143 | 14 | 74 | 60 |
| 15 | France | 0.6000 | 15 | 3 | (12) |
| 16 | Netherlands | 0.6000 | 15 | 16 | 1 |
| 17 | Belgium | 0.5857 | 17 | 28 | 11 |
| 18 | Kazakhstan | 0.5571 | 18 | 98 | 80 |
| 19 | Lithuania | 0.5286 | 19 | 20 | 1 |
| 20 | Slovenia | 0.5143 | 20 | 55 | 35 |

TABLE XI.    QUALITY OF E-PARTICIPATION WEBSITES OF SELECTED
COUNTRIES FOR 2010

| Range | Country | Score (%) | | | |
|-------|---------|-----------|---|---|---|
| | | E-information | E-consultation | E-decision making | Total |
| Over 60% | Republic of Korea | 87.50 | 78.79 | 75.00 | 78.95 |
| | Australia | 100.00 | 60.61 | 68.75 | 68.42 |
| | Kazakhstan | 87.50 | 66.67 | 62.50 | 68.42 |
| | Bahrain | 75.00 | 66.67 | 56.25 | 64.91 |
| | Spain | 75.00 | 63.64 | 37.50 | 57.89 |
| | Kyrgyzstan | 50.00 | 63.64 | 31.25 | 52.63 |
| | Mongolia | 62.50 | 54.55 | 43.75 | 52.63 |
| | Israel | 50.00 | 51.52 | 50.00 | 50.88 |
| | New Zealand | 50.00 | 54.55 | 43.75 | 50.88 |
| | United Kingdom of Great Britain | 50.00 | 60.61 | 31.25 | 50.88 |
| 30-60% | Japan | 87.50 | 39.39 | 50.00 | 49.12 |
| | United States of America | 50.00 | 54.55 | 37.50 | 49.12 |
| | Canada | 75.00 | 36.36 | 50.00 | 45.61 |
| | China | 37.50 | 39.39 | 62.50 | 45.61 |
| | Colombia | 75.00 | 39.39 | 43.75 | 45.61 |
| | Mexico | 87.50 | 51.52 | 12.50 | 45.61 |
| | Slovenia | 50.00 | 45.45 | 43.75 | 45.61 |
| | Chile | 75.00 | 39.39 | 31.25 | 42.11 |
| | Cyprus | 50.00 | 30.30 | 62.50 | 42.11 |
| | Estonia | 50.00 | 42.42 | 31.25 | 40.35 |
| | Singapore | 50.00 | 48.48 | 18.75 | 40.35 |
| Under 30% | Belarus | 37.50 | 33.33 | 18.75 | 29.82 |
| | France | 37.50 | 36.36 | 12.50 | 29.82 |
| | Netherlands | 75.00 | 18.18 | 31.25 | 29.82 |
| | Belgium | 62.50 | 12.12 | 43.75 | 28.07 |
| | Kenya | 37.50 | 33.33 | 12.50 | 28.07 |
| | Kuwait | 75.00 | 21.21 | 18.75 | 28.07 |
| | Turkey | 37.50 | 18.18 | 37.50 | 26.32 |

Source: Adapted from [7].

The brief discussion above reveals that the Bahrain government websites rank at a high level as e-service providers alongside with advanced and developed countries websites in the world. So far, Bahrain is the only country in the Gulf region that has managed to attain this result and serves as a very good example that can be followed by Saudi Arabia. Furthermore, the United Nations results showed a wide divergence between Bahrain and Saudi Arabia in terms of online service development, as Bahrain was ranked in 8[th] place worldwide compared to Saudi Arabia, which ranked 75[th] place.

The Saudi Arabian government should heed the web development progress of its ministries' websites. Moreover, it should treat this as a serious issue that impedes the current implementation of eGovernment, particularly in delivering eServices within the specified timeframe. Therefore, the Saudi Arabian government should speedily respond in handling this delay in the portal development of its ministries' websites and follow this up gradually with the top management or high authorities in the country in order to maintain a steady rate of web development for these government websites. Furthermore, engaging the public (citizens, residents and businesses) in decision-making is crucial and must be done to increase the level of transparency between the government and the public as well as to meet the public needs in easy, modern and effective ways.

References [25, 26] showed inactive role of the government supporting the growth of online activities in the country. They indicated that the government support is a critical key to promote online activities in the country as people and businesses in Saudi Arabia have tendency to feel more confident and secure with the online activities comes through the government or under its supervision.

## VI. Conclusion

Literature in the field of eGovernment implementation and specifically in the development of web portals for government agencies in developing countries, especially in Saudi Arabia, is still lacking. This motivates researchers to further explore this with the aim of bringing about the expected benefits of eGovernment implementation by pointing out the main barriers to eGovernment by analysing the web development progress of government websites. It is clear that some Saudi ministries have made progress in developing their websites to enhance the implementation of eGovernment applications.

However, these websites still need to provide more comprehensive online services that can adequately serve the residents/citizens. Providing eServices for car registration renewal via the traffic department website or for registering people in continuous educational programs through the education ministry website are examples of needed services. However, as the majority of Saudi ministry websites lack such services that would reinforce the concept of eGovernment among society, the Saudi government should pay close attention to the slow development of its government websites and take measures to remedy this immediately where applicable. This is particularly crucial as the majority of the selected websites in this study fall between Stages III and IV of the eGovernment stage model and some are providing poor online services as compared to Bahrain, which has achieved excellent results in this regard.

APPENDIX

| No. | Authority | Emerging presence | | | | | Enhanced presence | | | Interactive presence | | | | Transactional presence | | | Seamless | URL |
|---|---|---|---|---|---|---|---|---|---|---|---|---|---|---|---|---|---|---|
| | | agency name | Agency contacts | Agency address | Operation hour | General frequently asked questions | organizational news | publications | online policy | officials' email | post comment online | two-way communication | Download forms | e-form | e-payment | e-services | Full integration across organization | Website link |
| 1 | Saudi eGovernment national portal | ✓ | ✓ | ✓ | × | ✓ | ✓ | ✓ | ✓ | × | ✓ | ✓ | ✓ | × | × | × | × | http://www.saudi.gov.sa |
| 2 | Communication and Information Technology Commission | ✓ | ✓ | ✓ | ✓ | ✓ | ✓ | ✓ | ✓ | × | ✓ | ✓ | ✓ | ✓ | × | × | × | http://www.citc.gov.sa |
| 3 | Saudi Telecom | ✓ | ✓ | ✓ | × | ✓ | ✓ | ✓ | ✓ | × | ✓ | ✓ | × | ✓ | × | × | × | http://www.stc.com.sa |
| 4 | Ministry of Agriculture | ✓ | ✓ | ✓ | × | ✓ | ✓ | ✓ | ✓ | × | ✓ | ✓ | ✓ | ✓ | × | × | × | http://www.moa.gov.sa |
| 5 | Ministry of Civil Service | ✓ | ✓ | ✓ | × | ✓ | ✓ | ✓ | ✓ | × | ✓ | ✓ | ✓ | ✓ | × | ✓ | × | http://www.mcs.gov.sa |
| 6 | Ministry of Commerce and Industry | ✓ | ✓ | ✓ | × | × | ✓ | ✓ | ✓ | × | ✓ | ✓ | ✓ | ✓ | × | × | × | http://www.mci.gov.sa |
| 7 | Council of Saudi Chambers of Commerce and Industry | ✓ | ✓ | ✓ | × | ✓ | ✓ | ✓ | ✓ | ✓ | ✓ | ✓ | ✓ | ✓ | × | ✓ | × | http://www.saudichambers.org.sa |
| 8 | Ministry of Defense and Aviation | ✓ | ✓ | ✓ | × | ✓ | ✓ | ✓ | ✓ | × | ✓ | ✓ | ✓ | × | × | × | × | http://www.gaca.gov.sa |
| 9 | Ministry of Water and Electricity | ✓ | ✓ | ✓ | × | × | ✓ | ✓ | ✓ | × | ✓ | ✓ | ✓ | ✓ | × | ✓ | × | http://www.mowe.gov.sa/ |
| 10 | Saudi Ports Authority | ✓ | ✓ | ✓ | × | ✓ | ✓ | ✓ | × | × | ✓ | × | × | × | × | × | × | http://www.ports.gov.sa |
| 11 | Ministry of Interior Passport Authority | ✓ | ✓ | ✓ | × | ✓ | ✓ | ✓ | ✓ | × | ✓ | × | × | ✓ | × | ✓ | ✓ | http://www.gdp.gov.sa |
| 12 | Ministry of Health | ✓ | ✓ | ✓ | × | ✓ | ✓ | ✓ | ✓ | × | ✓ | ✓ | ✓ | ✓ | × | × | × | http://www.moh.gov.sa |
| 13 | Ministry of Education | ✓ | ✓ | ✓ | × | ✓ | ✓ | ✓ | ✓ | ✓ | ✓ | ✓ | ✓ | ✓ | × | × | × | http://www.moe.gov.sa |
| 14 | Ministry of Hajj | ✓ | × | ✓ | × | ✓ | × | ✓ | × | × | × | × | × | × | × | × | × | No official website |
| 15 | Ministry of Foreign Affairs | ✓ | ✓ | ✓ | × | ✓ | ✓ | ✓ | ✓ | × | ✓ | ✓ | ✓ | × | × | × | ✓ | http://www.mofa.gov.sa |
| 16 | Ministry of Higher Education | ✓ | ✓ | ✓ | × | ✓ | ✓ | ✓ | ✓ | × | ✓ | ✓ | ✓ | ✓ | × | ✓ | ✓ | http://www.mohe.gov.sa |
| 17 | Ministry of Communications and Information Technology (IT) | ✓ | ✓ | ✓ | × | ✓ | ✓ | ✓ | ✓ | × | ✓ | × | × | × | × | × | × | http://www.mcit.gov.sa |
| 18 | Ministry of Finance | ✓ | ✓ | ✓ | × | ✓ | ✓ | ✓ | ✓ | × | ✓ | ✓ | × | × | × | × | × | http://www.mof.gov.sa |
| 19 | Ministry of Justice | ✓ | × | × | × | ✓ | ✓ | ✓ | ✓ | × | ✓ | × | × | ✓ | × | × | × | http://www.moj.gov.sa |
| 20 | Ministry of Labor | ✓ | ✓ | ✓ | × | ✓ | ✓ | ✓ | ✓ | × | ✓ | × | × | ✓ | × | × | × | http://www.mol.gov.sa |
| 21 | Ministry of Municipalities and Rural Affairs | ✓ | ✓ | ✓ | × | ✓ | ✓ | ✓ | × | × | ✓ | × | × | × | × | × | × | http://www.momra.gov.sa/ |
| 22 | Ministry of Petroleum and Mineral Resources | ✓ | × | × | × | ✓ | ✓ | ✓ | ✓ | × | ✓ | × | × | × | × | × | × | http://www.mopm.gov.sa |
| 23 | Ministry of Economy and Planning | ✓ | ✓ | × | × | ✓ | ✓ | ✓ | ✓ | ✓ | ✓ | × | × | × | × | × | × | http://www.mep.gov.sa |
| 24 | Ministry of Islamic Affairs, Endowment, Dawa and Guidance | ✓ | ✓ | ✓ | × | ✓ | ✓ | ✓ | ✓ | × | ✓ | × | × | × | × | × | × | http://www.moia.gov.sa |
| 25 | Ministry of Social Affairs | ✓ | ✓ | ✓ | × | ✓ | ✓ | ✓ | × | × | ✓ | × | × | × | × | × | × | http://www.mosa.gov.sa |
| 26 | Ministry of Transport | ✓ | ✓ | ✓ | × | ✓ | ✓ | ✓ | ✓ | × | ✓ | × | ✓ | × | × | × | × | http://www.mot.gov.sa |
| 27 | Ministry of Interior | ✓ | ✓ | ✓ | × | ✓ | ✓ | ✓ | ✓ | × | ✓ | ✓ | × | ✓ | × | × | × | http://www.moi.gov.sa |
| 28 | General Presidency of Youth Welfare | × | × | × | × | × | × | × | × | × | × | × | × | × | × | × | × | http://www.gpyw.gov.sa  No access |